\documentclass[aps,twocolumn,amsfonts,amsmath,amssymb,prb]{revtex4}
\pdfoutput=1
\usepackage{graphicx}
\usepackage{graphics}
\newcommand*{\revrem}[1]{}
\usepackage{amsfonts}
\usepackage{amssymb}
\usepackage{bm}
\newcommand{\bmu}{\boldsymbol{\mu}}

\begin{document}

\title{Simulations of room temperature ionic liquids:\\ From polarizable to coarse-grained force fields}

\author{Mathieu Salanne$^{1,2}$}
\affiliation{$^{1}$~Sorbonne Universit\'es, UPMC Univ Paris 06, UMR 8234, PHENIX, F-75005 Paris, France. Tel: +33 144273265; E-mail: mathieu.salanne@upmc.fr}
\affiliation{$^{2}$~Maison de la Simulation, USR 3441, CEA - CNRS - INRIA - Universit\'e Paris-Sud - Universit\'e de Versailles, F-91191 Gif-sur-Yvette, France}

\begin{abstract}
Room temperature ionic liquids (RTILs) are solvent with unusual properties, which are difficult to characterize experimentally because of their intrinsic complexity (large number of atoms, strong Coulomb interactions). Molecular simulations have therefore been essential in our understanding of these systems. Depending on the target property and on the necessity to account for fine details of the molecular structure of the ions, a large range of simulation techniques are available. Here I focus on classical molecular dynamics, in which the level of complexity of the simulation, and therefore the computational cost, mostly depends on the force field. Depending on the representation of the ions, these are either classified as all-atom or coarse-grained. In addition, all-atom force fields may account for polarization effects if necessary. The most widely used methods for RTILs are described together with their main achievements and limitations. 
\end{abstract}

\maketitle


\section{Introduction}
RTILs are a particular class of solvents in which all the species are ionic. They have started to be widely investigated much more recently than conventional solvents such as water or organic liquids. As a consequence, unlike the latter which have generally been studied by theoretical approaches long after many experiments had been performed, RTILs are an interesting testing field for the predictive power of molecular simulations. In addition, most of the liquid state theories do not hold in these media due to the predominance of Coulombic interactions. A deep undestanding of the RTILs properties is nevertheless necessary for their use in a variety of applications. Indeed, although their industrial use is currently limited to the role of solvent for organic reactions, their good stability and their wide electrochemical window makes them excellent electrolytes, e.g. for energy storage devices.~\cite{armand2009a,macfarlane2014a} They are also able to stabilize new products such as nanoparticles with different properties than the ones produced in aqueous solvents.~\cite{pensado2011a,podgorsek2013a,mamusa2014a,mamusa2014b} There are thousands of possible cations/anions combinations, and the knowledge of the structure and transport properties, in the bulk or at interfaces, is mandatory for choosing the most appropriate one. Therefore, many studies involve the use of molecular simulations in order to understand and predict such properties. 

Although quantum chemistry methods are very useful for characterizing the nature of the interactions in RTILs,~\cite{hunt2006a,izgorodina2011a,zahn2013b,matthews2014a} these techniques are limited to small gas phase systems and they are not yet appropriate for the sampling of liquid properties. Most of the literature therefore involves molecular dynamics (MD) simulations. Interestingly, the outburst of ab initio molecular dynamics (AIMD), in which the atomic forces are calculated at each time step of the simulation using an electronic structure Density Functional Theory (DFT) calculation, has not been very visible in RTILs. There are two main explanations for this: Firstly, these liquids are strongly short-ranged ordered, each ion being surrounded by successive shells of alternating charged species. A relatively large number of ion pairs therefore have to be included in the simulation cells in order to avoid finite-size effects. Secondly, each molecule contains a large number of atoms and thus of electrons. The computational cost is therefore particularly large, compared to volvent-based electrolytes. The few AIMD studies have therefore focussed on the calculation of properties which are out of reach of classical MD, such as chemical reactivity~\cite{margulis2011a,xu2013a,kelemen2014a} or vibrational properties~\cite{thomas2014a}.  

\begin{figure}[h]
\centering
  \includegraphics[width=\columnwidth]{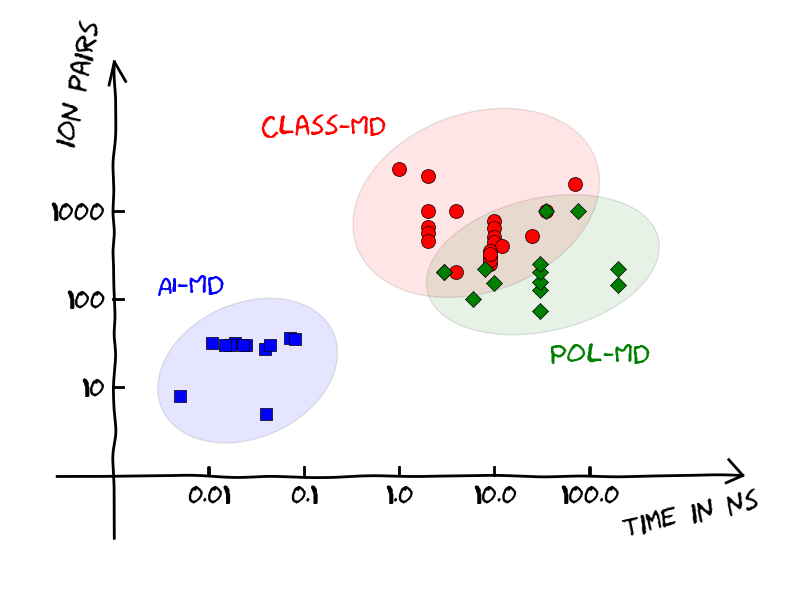}
  \caption{Non-exhaustive summary of the system sizes (number of ion pairs in the simulation cell) and the trajectory length simulated in a series of recent MD works. ``AI-MD'' stands for {Ab initio} MD, while ``Class-MD'' and ``Pol-MD'' respectively stand for classical MD involving non polarizable and polarizable force fields. Data extracted from references~\cite{xu2013a,kelemen2014a,thomas2014a,gabl2012a,schroder2012a,kashyap2012a,gontrani2012a,li2012b,liu2012b,salanne2012a,malberg2012a,pensado2012a,wendler2012a,shimizu2013a,weber2013a,wang2013e,ghatee2013b,mondal2014a,paredes2014a,tenney2014a,schmollngruber2014a,haskins2014a,starovoytov2014a,deoliveiracavalcante2014a}.}
  \label{fig:timings}
\end{figure}

The separation of time scales and length scales which can be studied by AIMD and classical MD is illustrated on figure \ref{fig:timings}, which reports data from a series of articles (from the period 2012--2014) for the number of ion pairs in the simulation cell with respect to the length of the trajectories in the production runs. It is not an exhaustive set of data and there are many biases which have to be kept in mind: To cite only one, AIMD is more sensitive to the number of electrons than to the number of ions. Nevertheless, it clearly shows that there are two orders of magnitude of difference in the size and time that can be studied with the two methods. Despite the development of very efficient codes, it is very unlikely that AIMD will be able to bridge this gap before long, so that classical MD will remain in future years the method of choice for many problems involving RTILs and requiring long simulation times.  

Although they share a similar methodology, i.e. the iterative integration of Newton's equation of motion over several millions of femtosecond timesteps, there exist many flavors of classical MD simulations. They are generally classified according to the nature of the interacting bodies and on the analytical expression for the interaction potential (or force field) which acts between them. For the former, the most general way to handle molecular species is to assign one interaction site to each atom (all atom force fields). This approach may be simplified by coarse-graining, for example by gathering the hydrogen atoms with the atoms to which they are linked (united atoms force fields), or even further by including a larger number of atoms into one grain (coarse-grained force fields). There is an even larger range of choices for the analytical expression of the interaction potential. In this article, the most widely used methods for RTILs will be described. The following key points will be addressed: In each case, the general shape of the force field will be provided, together with some general methods for obtaining the corresponding parameters. Then the capabilities and limitations of each method will be discussed. 

\section{Non-polarizable all atom force fields}

As in many other fields, most of the classical MD simulations of RTILs involve non-polarizable force fields and they generally include all the atoms from the molecules as interaction sites. Since ionic liquids are made of functional groups which are present in many conventional organic molecules, most of the force fields have been parameterized following the existing families, such as AMBER or OPLS, in which the total interaction potential is composed of bonded and non-bonded terms,

\begin{equation}
E_{\rm total}=E_{\rm bonds}+E_{\rm angles}+E_{\rm dihedrals}+E_{\rm non-bonded}
\end{equation}

\noindent where the bonded terms $E_{\rm bonds}$, $E_{\rm angles}$ and $E_{\rm dihedrals}$ take the following forms:
\begin{eqnarray}
E_{\rm bonds} & = & \sum_{ij}^{\rm bonds} \frac{k^r_{ij}}{2}\left(r_{ij}-r^0_{ij}\right)^2 \label{eq:bond}\\
E_{\rm angles} & = & \sum_{ijk}^{\rm angles} \frac{k^\theta_{ijk}}{2}\left(\theta_{ijk}-\theta^0_{ijk}\right)^2 \label{eq:angle} \ \\
E_{\rm dihedrals} & = & \sum_{ijkl}^{\rm dihedrals} \sum_{m=1}^4 \frac{V^m_{ijkl}}{2}\left[1+(-1)^m\cos{m \phi_{ijkl}} \right] \label{eq:dihedral}
\end{eqnarray}

\noindent The sums respectively run over the full set of instantaneous bonds $r_{ij}$, angles $\theta_{ijk}$ and dihedrals $\phi_{ijkl}$ which must be defined prior to the simulations (i.e. the bonds are not allowed to break or form during the simulation). The quantities $k^r_{ij}$, $r^0_{ij}$, $k^\theta_{ijk}$, $\theta^0_{ijk}$  and $V^m_{ijkl}$ are parameters of the force field. 

The non-bonded term is in principle composed of three contributions

\begin{equation}
E_{\rm non-bonded}=E_{\rm repulsion} + E_{\rm dispersion} + E_{\rm electrostatic}
\end{equation}

but  the repulsion and dispersion terms are gathered into a Lennard-Jones potential:

\begin{eqnarray}
E_{\rm Lennard-Jones}&=&E_{\rm repulsion} + E_{\rm dispersion} \label{eq:lj}\\
                     &=&\sum_{i}\sum_{j>i}4\epsilon_{ij}\left[\left(\frac{\sigma_{ij}}{r_{ij}}\right)^{12}-\left(\frac{\sigma_{ij}}{r_{ij}}\right)^{6}\right] \nonumber
\end{eqnarray}

Finally, the electrostatic interactions are calculated with the usual Coulomb potential: 

\begin{equation}
E_{\rm electrostatic}=E_{\rm Coulomb}=\frac{1}{4\pi\epsilon_0}\sum_{i}\sum_{j>i}\frac{q_iq_j}{r_{ij}} \label{eq:coulomb}
\end{equation}

Each atom type $i$ is therefore characterized by three parameters only, $\epsilon_{i}$, $\sigma_{i}$ and $q_i$; the parameters of the Lennard-Jones potential are then derived using the Lorentz-Berthelot mixing rules:
\begin{eqnarray}
\epsilon_{ij}&=&\sqrt{\epsilon_{i}\epsilon_{j}}\\
\sigma_{ij}&=&\frac{1}{2}\left(\sigma_{i}+\sigma_{j}\right)
\end{eqnarray}

\begin{figure}[h]
\centering
  \includegraphics[width=\columnwidth]{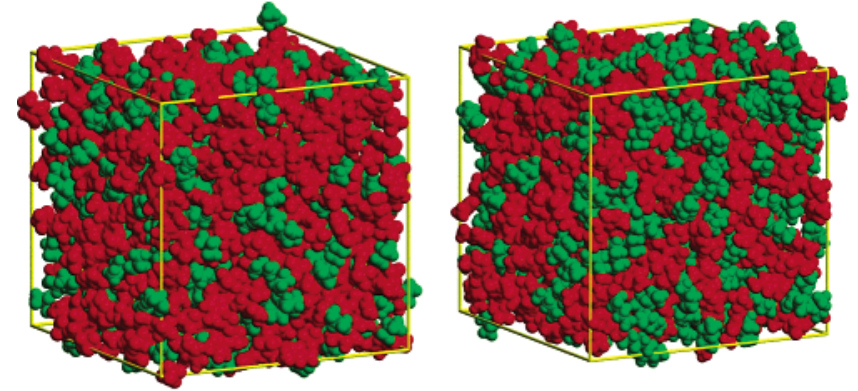}
  \caption{Snapshots of simulation boxes, using a coloring code which identifies the charged (red) and nonpolar (green) domains of RTILs. Left: BMIM-PF$_6$, right: HMIM-PF$_6$. Reproduced with permission from ref. \cite{canongialopes2006a}. Copyright 2006 American Chemical Society.}
  \label{fig:nanostructure}
\end{figure}

The most famous force field for RTILs is the CL\&P~\cite{canongialopes2012a}, which was built based on the OPLS functional form. It was parameterized for a large set of ionic liquids compounds, including the imidazolium, N-alkyl-pyridinium, tetraalkylammonium, N,N-pyrrolidinium and tetralalkylphosphonium families for cations, and the chloride, bromide, triflate, bis(sulfonyl)imide, alkylsulfate, alkylsulfonate, phosphate, nitrate and dicyanimide families for anions.~\cite{canongialopes2004a,canongialopes2004b,canongialopes2004c,canongialopes2006b,canongialopes2008a,shimizu2010a,canongialopes2012a} The bonded and Lennard-Jones parameters were generally taken form the OPLS set~\cite{jorgensen1996a} when available and led to accurate results, but some of them had to be determined or reparameterized for consistency. This was generally done by reproducing the molecular geometries and the torsion energy profiles calculated for isolated molecules using quantum chemistry methods (at the MP2 level). The partial charges were optimized to reproduce the electrostatic field generated by the molecule, using the CHelpG method.~\cite{breneman1990a}

The first simulation studies of RTILs have focussed on the understanding of their structure~\cite{price2001a,deandrade2002a,lyndenbell2007a}. They showed that, like in conventional molten salts, it is dominated by a combination of short-range repulsion and Coulomb ordering effects. Around a given ion, the first neighbour shell consists entirely of oppositely charged species; this ordering automatically transfers up to several neighbour shells. The short-range structure is somewhat more complex: In particular, the three-dimensional arrangement of the anions around the cationic molecules strongly depends on the molecular shape of the latter. For example, weak hydrogen bonds form between the protons of imidazolium rings and the most electronegative atoms of the anions.~\cite{fumino2014a,salanne2006b}   

But the structure of RTILs is also characterized by strong intermediate range ordering, which can be detected by the presence of a pre-peak or a shoulder in the low wavevector part of X-ray and neutron diffractograms. Ribeiro {\it et al.} showed that both the position and the intensity of this prepeak are sensitive to the length of alkyl chains of imidazolium cations.~\cite{urahata2004a} The presence of an additional ordering was then clearly characterized by Canongia Lopes and P\'adua. By simulating various RTILs using the CL\&P and distiguishing the highly charged, ``polar'' regions of the ions from the ``unpolar'' ones, they have put in evidence the presence of nanostructured domains.~\cite{canongialopes2006a} Their famous pictures of RTILs are reproduced on figure \ref{fig:nanostructure} for two different liquids, namely the 1-butyl-3-methylimidazolium hexafluorophosphate (BMIM-PF$_6$) and the 1-hexyl-3-methylimidazolim hexafluorophosphate (HMIM-PF$_6$). This important finding is now commonly used to interpret the solvation properties of RTILs towards many species, such as nitrate ions~\cite{figueiredo2012a,hayes2014a} or acidic (SO$_2$, CO$_2$) gases.~\cite{firaha2014a,morganti2014a} It was also extended to interpret the structure of confined ionic liquids as probed in Atomic Force Microscopy or Surface Force Apparatus experiments~\cite{perkin2010a,perkin2011a,perkin2012a,smith2013a,li2014d} and in MD simulations~\cite{duarte2014a}. The existence of this intermediate range ordering has a strong implication on the lubrication properties of RTILs.~\cite{smith2014a} 



\section{Polarizable all atom force fields}

\begin{figure}[h]
\centering
  \includegraphics[width=\columnwidth]{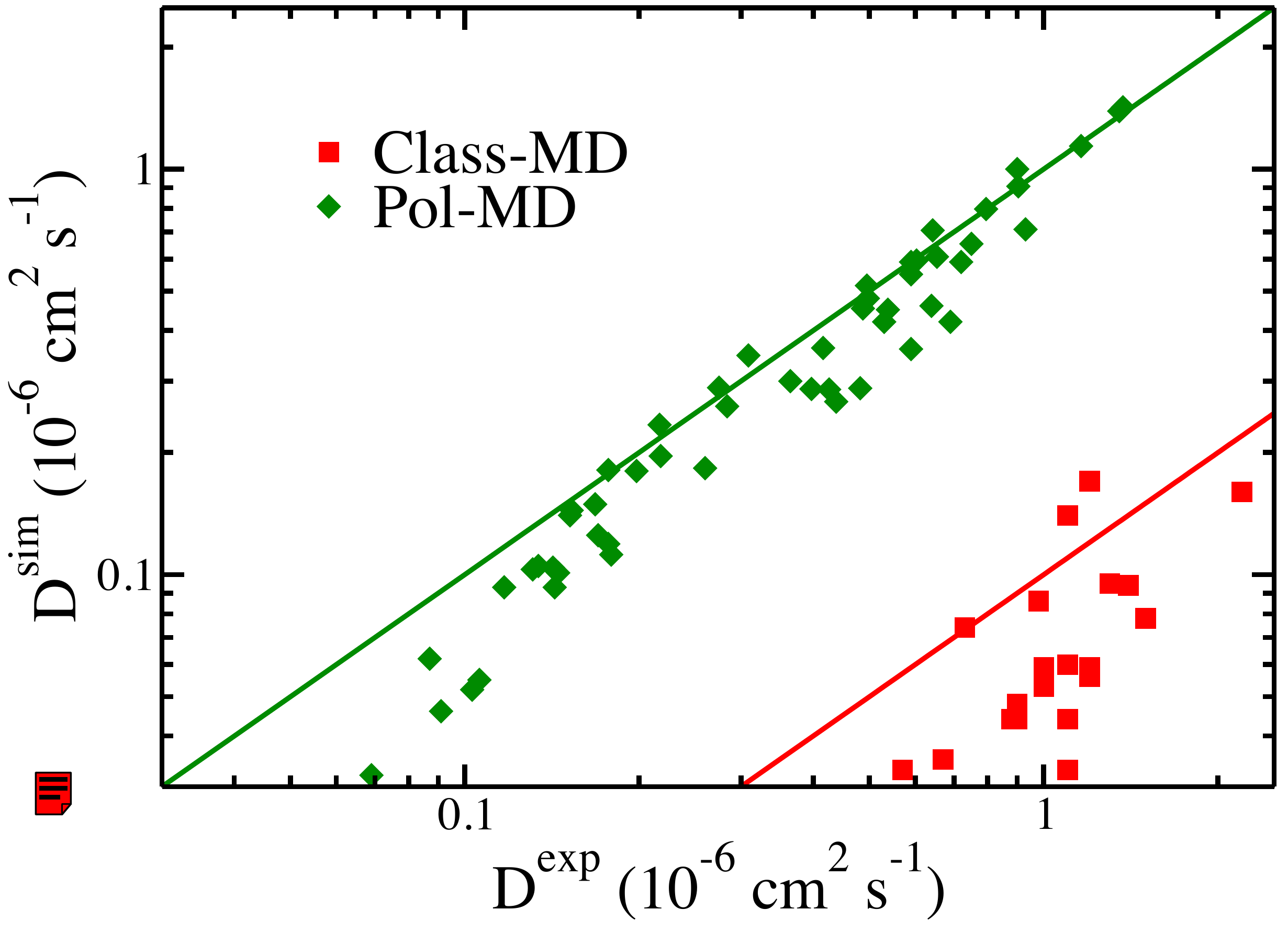}
  \caption{Comparison of the simulated and experimental diffusion coefficients for a series of ionic liquids, using polarizable (Pol-MD) and non-polarizable (Class-MD) force fields. The corresponding data are respectively extracted from references \cite{borodin2009a} and \cite{tsuzuki2009a}.}
  \label{fig:diffusion}
\end{figure}

Although the CL\&P (and related) force fields have led to a deep understanding of the structure of RTILs, they generally fail to predict the transport properties. Indeed, they yield viscosities which are too high by one order of magnitude on average, while diffusion coefficients and conductivities are underestimated accordingly. This discrepancy is shown on figure \ref{fig:diffusion}, where the simulated diffusion coefficients are compared to the experimental ones (as measured by pulse field gradient NMR) for a series of common ionic liquids (data extracted from reference \cite{tsuzuki2009a}). On this plot, the red line corresponds to an underestimation by a factor of 10. We observe that not only are all the values largely underestimated, but also the data are largely dispersed and there is not a clear correlation between them. This means that even comparing the simulated diffusivities for two ionic liquids may not give a qualitatively meaningful result, and that any dynamic information extracted from such a simulation should be taken with caution.

\begin{figure}[h]
\centering
  \includegraphics[width=\columnwidth]{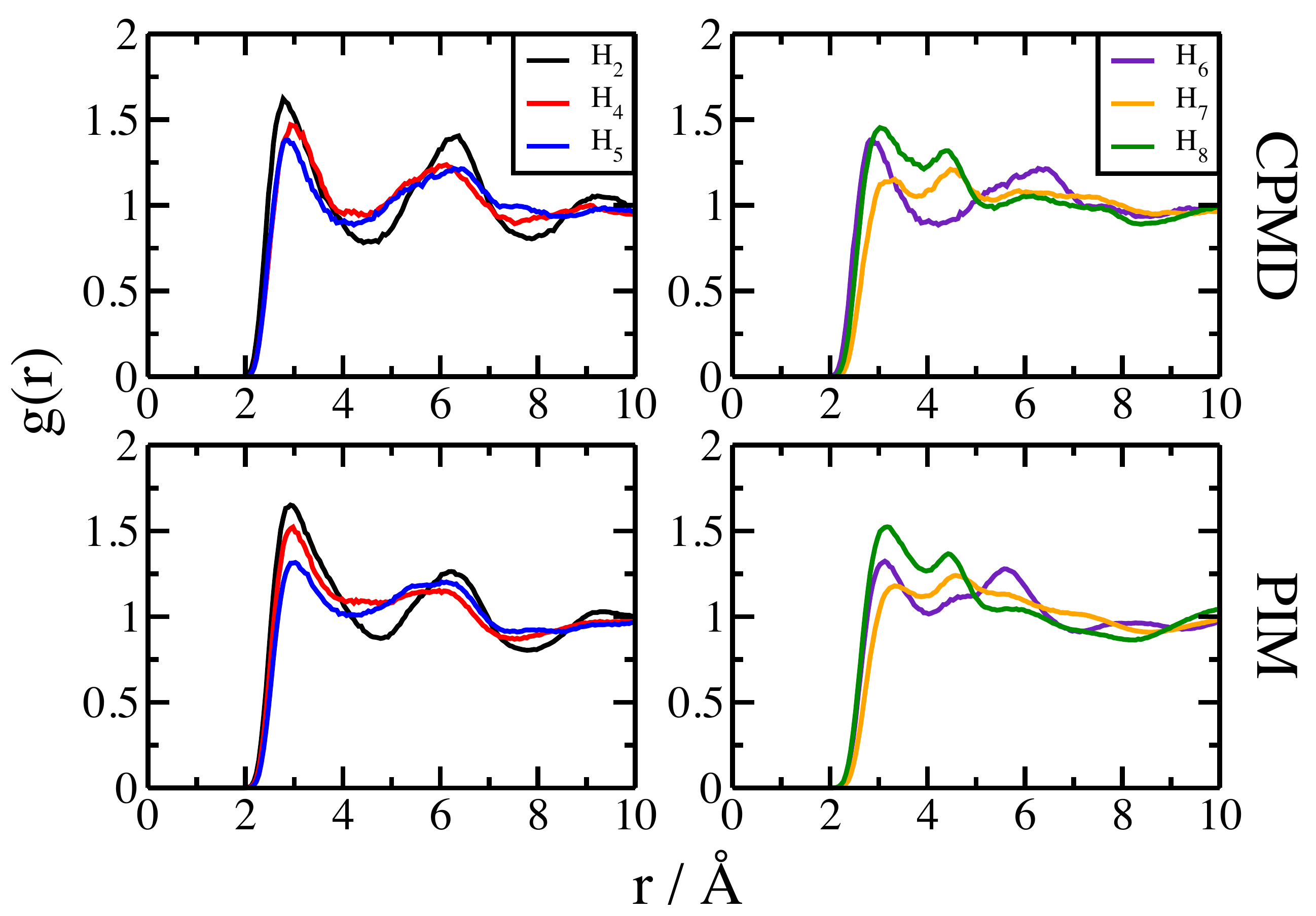}
  \caption{Radial distribution functions between the chloride anions and the protons from the EMIM$^+$ cations in the EMIM-AlCl$_4$ RTIL at 298~K. Top: AIMD results (labelled ``CPMD'' because they were obtained using the Car-Parrinello method); bottom: classical MD results with a polarizable force field (PIM). Left: Protons from the imidazolium ring. Right: Protons from the alkyl chains. Reproduced with permission from ref. \cite{salanne2012a}.}
  \label{fig:compcpmd}
\end{figure}

The solution to overcome this difficulty was to include polarization effects explicitly in the force field. Indeed, since the pioneering study by Yan {\it et al.},~\cite{yan2004a} there have been numerous examples showing an increase of the diffusivity when using polarizable force fields. Nevertheless, these comparisons are often difficult to interpret because one cannot generally add a polarization term on top of an existing force field. It is necessary to re-parameterize part or all of the other terms. In a recent work, we have focussed on the mixtures of 1-ethyl-3-methylimidazolim chloride (EMIM-Cl) with aluminium chloride (AlCl$_3$).~\cite{salanne2012a} At equimolar proportions, all the chloride ions are linked with aluminium ones through very strong ionic bonds, thus forming the ionic liquid EMIM-AlCl$_4$. To simulate it, we have used the CL\&P force field for all the interactions (both bonded and non-bonded) involving EMIM$^+$ cations only. The Polarizable Ion Model (PIM)~\cite{salanne2012b} was used for all the other interactions. In this model the repulsion and dispersion terms are described as 
\begin{eqnarray}
E^{\rm PIM}_{\rm vdw}&=&E_{\rm repulsion}+E_{\rm dispersion}\\
&=&\sum_i \sum_{j > i} \left( B_{ij}{\rm e}^{-a_{ij} r_{ij}}  -f^{ij}_6(r_{ij})\frac{C_6^{ij}}{r_{ij}^6}-f^{ij}_8(r_{ij})\frac{C_8^{ij}}{r_{ij}^8} \right)\nonumber
\end{eqnarray}

\noindent where the $C_6^{ij}$ and $C_8^{ij}$ are the dipole-dipole and dipole-quadrupole dispersion coefficients; $f_n$ are Tang-Toennies dispersion damping functions~\cite{tang1984a} describing the short-range penetration correction to the asymptotic multipole expansion of dispersion. These functions take the form

\begin{equation}
f_n^{ij}(r_{ij})=1-{\rm e}^{-b_n^{ij}r^{ij}}\sum_{k=0}^n\frac{(b_n^{ij}r_{ij})^k}{k!}
\end{equation}

The most important changes concern the electrostatic term, which now includes charge-dipole and dipole-dipole interaction:

\begin{eqnarray}
\label{eq:polarization}
E^{\rm PIM}_{\rm electrostatic}&=&E_{\rm Coulomb}+E^{\rm PIM}_{\rm polarization}\\
&=&\frac{1}{4\pi\epsilon_0}\sum_{i}\sum_{j>i}\left(\frac{q_iq_j}{r_{ij}}-  g^{ij}(r^{ij})q_i{\mathbb T}_1^{ij}\bmu_j \right. \nonumber\\ 
& & \left. + g^{ji}(r^{ij})q_j{\mathbb T}_1^{ij}\bmu_i-\bmu_i{\mathbb T}_2^{ij} \bmu_j \right) \nonumber\\ 
& & +\sum_i\frac{1}{2\alpha_i}\bmu_i^2 \nonumber
\end{eqnarray}

\noindent where $\bmu_i$ is the induced dipole of ion $i$, while ${\mathbb T}_1$ and ${\mathbb T}_2$ are the charge-dipole and dipole-dipole interaction tensors defined by
\begin{eqnarray}
{\mathbb T}_1^{ij}&=& {\boldsymbol \nabla}_i \frac{1}{r_{ij}} = -\frac{{\bf r}_{ij}}{r^3_{ij}}  \\
{\mathbb T}_2^{ij}&=& {\boldsymbol \nabla}_i {\mathbb T}_1^{ij}=\frac{3{\bf r}_{ij}\times{\bf r}_{ij}}{r_{ij}^5}-\frac{1}{r_{ij}^3}{\mathbb I}
\end{eqnarray}
\noindent and $\alpha_i$ is the polarizability of ion $i$, which is assumed to be isotropic. The last term in equation \ref{eq:polarization} accounts for the energy cost of deforming the electron densities of the ions to create the induced dipoles. Note that Tang-Toennies functions are also included to account for the short-range damping effects on the charge-dipole interactions:

\begin{equation}
g^{ij}(r_{ij})=1-c_{ij}{\rm e}^{-b_{ij}r_{ij}}\sum_{k=0}^4\frac{(b_{ij}r_{ij})^k}{k!}
\end{equation}

The parameters in the polarizable potentials are then optimized by matching the dipole and forces obtained using the potential to the ones calculated by DFT on a series of representative configurations of liquid chloroaluminates,~\cite{kirchner2006a,kirchner2007a} following the method described in references \cite{aguado2003b} and \cite{salanne2012b}. The force field was then validated in two steps: Firstly, the structure was compared to that calculated with an AIMD simulation. Various radial distribution functions provided by the two methods (the AIMD one is labelled ``CPMD'' since it was obtained using the Car-Parrinello method) are shown on figure \ref{fig:compcpmd}. An excellent agreement is observed for all the atoms. We note that the agreement is much better than the one obtained by Youngs {\it et al.} who have tried to fit a non-polarizable force field for the ionic liquid dimethylimidazolium chloride using a similar force-fitting technique.~\cite{youngs2006a} Although they were able to reproduce the peak positions better than with previous classical MD potentials, the intensities of the first peaks where largely overestimated (especially in the case of the ring protons). The over-structuring obtained by these authors could partly be cancelled by multiplying all the $\epsilon_i$ parameters by a factor of 2, at the price of a worse reproduction of the initial set of DFT forces. Taking into account anion polarization effects leads to strong improvements in the description of the structure of the RTILs.

Secondly, the dynamical properties  determined with the PIM force field for EMIM-AlCl$_4$ were compared to the available experimental data. An excellent agreement was found for the diffusion coefficients, the viscosity and the electrical conductivity,~\cite{salanne2012a} confirming that the polarization term is compulsory for studying transport properties of RTILs. This is remarkable since the force field was parametrized using DFT calculations only, no experimental information was included in the fitting procedure. Another interesting aspect of our PIM is that it was made consistent with our previous work on molten salts,~\cite{salanne2011a} i.e. the chloroaluminate anion is represented by Al$^{3+}$ and Cl$^-$ ions, which are bound together only with by strong electrostatic interactions. It could therefore be used for example for studying non-stoichiometric mixtures of EMIM-Cl and AlCl$_3$, something which would not be possible with a conventional non-polarizable force field. 

Polarizable force fields have also been developped for a series of RTILs by Borodin, containing 1-methyl-3-alkylimidazolium, 1-alkyl-2-methyl-3-alkylimidazolium, N-methyl-N-alkylpyrrolidinium, N-alkylpyridinium, N-alkyl-N-alkylpiperidinium, N-alkyl-N-alkylmorpholinium, tetraalkylammonium, tetraalkylphosphonium, N-methyl-N-oligoetherpyrrolidinium cations and BF$_4^-$, CF$_3$BF$_3^-$, CH$_3$BF$_3^-$, CF$_3$SO$_3^-$, PF$_6^-$, dicyanamide, tricyanomethanide, tetracyanoborate, bis(trifluoromethane sulfonyl)imide (Ntf$_2^-$ or TFSI$^-$), bis(fluorosulfonyl)imide (FSI$^-$) and nitrate anions.~\cite{borodin2009a} The bonded terms have a similar analytical form as in the CL\&P while the non-bonded terms resemble the PIM ones. For the latter, the only difference is that the Tang-Toennies functions are replaced by a strongly repulsive term at (very) short-range for the dispersion, and through the use of Thole screening functions~\cite{thole1981a} for the polarization. This polarizable force field was parameterized by combining quantum chemistry data (for the bonded and electrostatic terms) and experimental results for the repulsion and dispersion terms. For the latter, in addition to the densities, the diffusion coefficients were used in the fitting procedure. As shown in figure \ref{fig:diffusion}, they are very well reproduced by the force field:  only small discrepancies are observed for the lowest values. Such an agreement is of course partly due to the parameterization procedure, but it is likely that it would not be possible to obtain it using a non-polarizable force field. Another example of excellent reproduction of the dynamical properties was obtained by Choi {\it et al.} for the BMIM-BF$_4$ ionic liquid. Their polarizable interaction potential was entirely parametrized from ab initio calculations, using the symmetry-adapted perturbation theory (SAPT). The diffusion coefficients and electrical conductivities calculated with this potential are in close agreement with the experimental data at several temperatures.~\cite{choi2014b}

\section{On the use of reduced charges in non polarizable force fields}

It is worth noting that in recent years many papers have discussed the opportunity to use non-polarizable force fields, but with reduced charges. Indeed, although the atomic charge is not a well-defined quantity in quantum chemistry or DFT calculations, it is possible to derive some values using a variety of approaches. The most popular ones are the CHelpG,~\cite{breneman1990a} RESP,~\cite{bayly1993a} Bader analysis~\cite{bader1990a} and Bl\"ochl~\cite{blochl1995a} methods. All these methods, when used on ionic liquids with no constraint on the total charge of the ions, yield values smaller than the formal charge.~\cite{schmidt2010a,zhang2012c,holloczki2014a} There are two interpretations: This may be either due to polarization or to charge transfer effects. Discriminating between these effects is a tricky issue, and it is rendered even more difficult by the fact that all the charge derivation methods provide different values for the same configuration and that even a given method yields different sets of charges depending on the conditions of the calculation (i.e. choice of the grid points for calculating the electrostatic potential, different conformations, gas-phase or condensed phase calculation, etc). Nevertheless the success of the PIM for  a large variety of inorganic ionic materials~\cite{salanne2011c} and of polarizable force fields for RTILs seem to indicate that charge transfer is negligible in these systems.  An advantage of the SAPT approach by Choi {\it et al.} is that the induction energy, which contains both the polarization and the charge transfer terms, is calculated with ab initio techniques. In their study, they concluded that polarization is the dominant energy  based on the fact that it is extremely well-reproduced by their force field which includes polarization only (the ab initio calculation was performed on 1300 BMIM-BF$_4$ complexes extracted from the liquid).~\cite{choi2014b} 

Using reduced charges should therefore be seen as an effective way to account for polarization effects only. A rationale for this approach was provided by Leontyev and Stuchebrukhov, who explain the need for charge-scaling in non-polarizable force fields by a neglect of the electronic solvation energy.~\cite{leontyev2011a} They therefore suggest to screen the Coulomb interactions by a factor $1-1/\epsilon_{\it el}$, where $\epsilon_{\it el}$ is the electronic (high-frequency) dielectric constant, which can be determined from experimental measures of the refractive index~\cite{bica2013a} or from a direct computation of the molecular polarizabilities using DFT.~\cite{salanne2008e} In order to check how reliable this simplified approach is, Schr\"oder performed an extensive comparison of the two methods. In his study, he progressively ``switched on'' the polarization effect (using Drude oscillators instead of the explicit induced dipoles discussed here, but this should not impact much the conclusions~\cite{schmollngruber2014b}), either by increasing the polarizability or by scaling down the charges with a scaling factor $f$ ranging from 0 to unity (i.e. the simulation with $f=0$ corresponds to the non-polarizable force field, while $f=1$ corresponds to the normal polarizable force field or to the non-polarizable force field with scaled charges).~\cite{schroder2012a} He showed that although the reduced charge force field was not able to reproduce the correct dipole distribution, leading to substantial deviations for the mean rotational relaxation time, the diffusion coefficients and the electrical conductivity were qualitatively correct. In conclusion, the reduced charge method can be considered appropriate if one is interested in recovering the correct order of magnitude for the dynamical properties, but it is important to keep in mind that the local relaxation mechanisms may not be accurate. 

As discussed in the introduction, an important aspect of a simulation is its computational cost. Including explicit dipoles increases the simulation times by a factor of 5 to 10 for a given number of atoms, which explains why they are often discarded. Nevertheless, we observe on figure \ref{fig:timings} that this order of magnitude in the cost does not seem to be reflected in practice. Of course, the data shown here is far from being exhaustive and there are many possible biases (for example, it is possible that the groups who have access to larger computational ressources  will tend to use a more expensive method for tackling a given problem), but the expected difference of one order of magnitude is not clearly observed. Indeed, although larger systems are generally simulated by non-polarizable force fields, the longer trajectories have been accumulated using polarizable ones. These observations may be explained by the fact that the polarizable force fields are generally used for calculating transport properties while the non-polarizable ones focus on the structure. Note also that in classical MD the simulation time scales as $N^2$ where $N$ is the number of interacting sites, while extending a trajectory over time leads to a linear increase of the simulation time. As a consequence, it is more efficient to do a longer simulation time on a smaller simulation cell for studying the transport properties with polarizable force fields. Nevertheless figure \ref{fig:timings} shows that the continuous increase in the computational ressources available, together with the efficient parallelization of MD codes, has allowed polarizable force fields to become a viable alternative for studying the physical properties of RTILs. 

\section{Coarse-grained force fields}

Coarse-grained force fields are used when the computational cost needs to be strongly reduced. For example, many problems in material science in which ionic liquids are used involve complex interfaces. In energy storage devices such as supercapacitors or batteries, RTILs play the role of the electrolyte, in which the transport of the charge occurs via the migration of ionic species from one electrode to the other.~\cite{armand2009a,fedorov2014a} Understanding how they operate then necessitates very large simulation cells with explicit electrodes. This limits the number of systems that can be studied and prevents systematic comparisons. Here we will focus on the topic of supercapacitors, which has attracted a lot of attention from the RTIL modelling community in the last years. Indeed, the energy storage does not imply any (electro)chemical reaction, but rather the reversible adsorption of the ions on porous carbon electrodes, so that this problem can be tackled using classical molecular simulation methods.  A first understanding of the adsorption behavior of RTILs on electrodes has been provided by analytical theories.~\cite{kornyshev2007a} Then the first simulations have been performed using simplified models of RTILs. For example, Fedorov {\it et al.} devised some generic rules on the behavior of RTILs at the surface of planar electrodes, with an emphasis on the effects of the ionic sizes, of the presence/absence of neutral groups (representing the alkyl chains) on the cations and of the voltage.~\cite{fedorov2008a,georgi2010a,ivanistsev2014a,ivanistsev2014b} Nevertheless it is worth underlining that these simplified models cannot really be considered as coarse-grained force fields because there is no mapping between them and real ionic liquids, so that it is not possible to use them directly for the study of a specific system.

\begin{figure}[h]
\centering
  \includegraphics[width=\columnwidth]{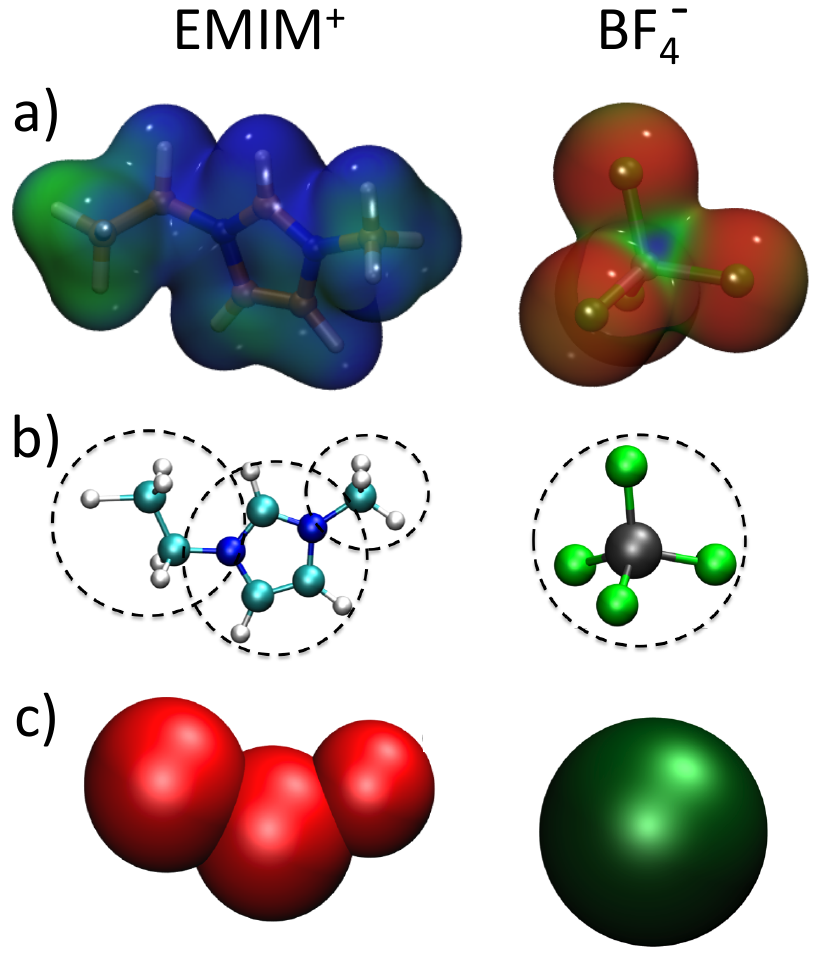}
  \caption{Coarse-graining of the EMIM$^+$-BF$_4^-$ ionic liquid. a) Electron density isosurface mapped with the corresponding electrostatic potential. b) All-atom representation. c) Coarse-grained representation. }
  \label{fig:coarsegraining}
\end{figure}

The two main challenges for simulating {\it realistic} supercapacitors are i) the size of the system and ii) the handling of constant potential conditions. Indeed, they involve nanoporous carbon materials,~\cite{simon2008a,frackowiak2013a} which can only be modelled with complex structures. There are several methods for maintaining those at constant potential.~\cite{merlet2013c} In our model,~\cite{siepmann1995a,reed2007a,limmer2013a} the local charges on the atoms of the electrode vary dynamically in response to the electrical potential caused by the ions and molecules in the electrolyte. They are therefore calculated at each time step of the simulation using a self-consistent approach similar to the one used for calculating the induced dipoles in polarizable force fields, so that the computational cost associated with such simulations is very high. For these reasons, simulations at constant applied potential and using all atom force fields have mostly been limited to electrodes with simpler geometries and the computational cost is often reduced by using multiple time step algorithms, i.e. the values of the electrode charges are not calculated at each step of the simulation.~\cite{vatamanu2010b,vatamanu2013a,vatamanu2014a}

In our recent work we have focussed on supercapacitors involving carbide-derived carbon electrodes and RTIL electrolytes, which present very promising experimental performances.~\cite{simon2008a,largeot2008a} The carbon structures, obtained by Palmer {\it et al.},~\cite{palmer2010a} are made of around 4000 atoms and have pore size distributions and accessible surfaces similar to the real devices.~\cite{dash2006a} Due to the large number of electrode atoms, it was not possible to use an all atom force field for the RTIL, so that we resorted to a coarse-grained reprensentation of the ions instead. Experimental works have shown that the most important effect on the capacitance is due to the matching between the pore size and the ion size~\cite{largeot2008a,chmiola2006a}, so this approximation should not affect the results from the qualitative point of view. The possibilities to describe a RTIL using a coarse-grained approach are very large.  Prior to the choice of the functional form for the non-bonded terms (and to the determination of the corresponding parameters), it is necessary to establish which atoms will be put together in a grain and whether the ``bonds'' should be held rigid or treated with bonded terms similar to those of equations \ref{eq:bond}, \ref{eq:angle} and \ref{eq:dihedral}. In a first step, simple RTILs based on BMIM$^+$ and EMIM$^+$ cations and on PF$_6^-$ and BF$_4^-$ anions were considered. As can be seen on the panel a) of figure \ref{fig:coarsegraining}, the electrostatic potential (extracted from a DFT calculation) of these molecules is rather smooth, so that it is possible to pass from the all-atom representation of panel b) to the coarse-grained one of panel c). Such a coarse-graining approach was proposed for BMIM-PF$_6$ by Roy {\it et al.}~\cite{roy2009a} In their representation, the BMIM$^+$ cation is represented by three interaction sites (one for the imidazolium ring, one for the methyl group and one for the butyl group) while the PF$_6^-$ anion is replaced by a single site. The cation is kept rigid during the simulation, its geometry is therefore chosen according to its average conformation in the liquid. The non-bonded interaction are calculated using Lennard-Jones (equation \ref{eq:lj}) and Coulomb (equation \ref{eq:coulomb}) potentials. The corresponding parameters ($\epsilon_i$, $\sigma_i$ and $q_i$) do not have an atomic character anymore, so that they are less transferable. Starting from an initial guess based on an all-atom force field, they are chosen in order to match as closely as possible the experimental density and diffusion coefficients, so that they are purely empirical. In a first step, unit charges were attributed to the ions.~\cite{roy2009a} As for all atom non polarizable force field, this led to understimated transport properties. The use of reduced charges~\cite{roy2010a} then corrected this discrepancy. In the case of a coarse-grained model, this choice is fully justified since there should be a strong screening from the whole grain associated to each charge. This coarse-grained force field was then validated for interfacial properties (surface tension)~\cite{merlet2011a} and extended to EMIM$^+$ and BF$_4^-$ ions~\cite{merlet2012b}.

\begin{figure}[h]
\centering
  \includegraphics[width=\columnwidth]{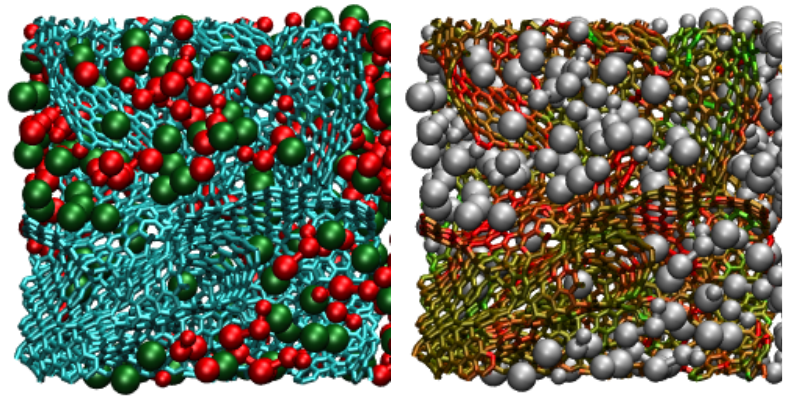}
  \caption{Instantaneous structure of the BMIM-PF$_6$ ionic liquid adsorbed in a nanoporous carbon electrode held at 0.5~V. Left: Position of the atoms (turquoise: carbon, green: PF$_6^-$, red: BMIM$^+$). Right: Local charge of the various carbon atoms (green: negative charge, red: positive charge and yellow: null charge); the atoms of the ionic liquids are shown in gray. }
  \label{fig:screening}
\end{figure}

Using this potential, we have been able to obtain much larger capacitances in nanoporous carbons than with planar electrodes, in agreement with the experimental results.~\cite{merlet2012a} We have shown that the electrode is wetted by the electrolyte at null potential and that the charging process involves the exchange of ions with the bulk electrolyte without substantially changing the volume of liquid inside the electrode. This exchange is accompanied by a partial decrease of the coordination number of the ions rendered possible by the charge compensation by the electrode, which underlines the importance of the electrostatic interactions on the charge storage mechanism. The screening of the Coulomb interaction by the metal at non-zero voltages is even at the origin of the formation of a ``superionic'' state,~\cite{kondrat2011a,kondrat2011b} in which two cations or two anions are contiguous as can be seen on figure \ref{fig:screening}, a situation which is extremely unfavorable in a bulk RTIL. Then we demonstrated the key role of the local structure. In qualitative agreement with recent NMR studies,~\cite{wang2013d,forse2013a,deschamps2013a} our simulations could put in evidence that the adsorption of the ions occurs on various sites in a way which depends on the applied potential.~\cite{merlet2013d} Four different types of sites were proposed in the case of CDCs: Depending on the number of carbon atoms in the vicinity of the ions, we have distinguished edge sites (concave curvature), plane sites (graphene sheet-like structure), hollow sites (convex curvature) and pocket sites (inside a subnanometre carbon pore). More recently, we could exploit the fact that these coarse-grained force fields were adjusted in order to reproduce well the diffusion coefficients for studying the charging dynamics of these supercapacitors.~\cite{pean2014a} Characteristic charging times of approximately 1 second were determined by extrapolating the MD results to a macroscopic electrode, here again in good agreement with the available experimental data.  

Although very efficient, these coarse-grained force fields suffer from a lack of transferability. It is probably necessary to reparameterize them and to redefine the geometry of the ions when passing from a RTIL to another. If more complex ionic liquids are simulated in the future, it will probably be necessary to use the effective force coarse-graining approach proposed by Voth {\it et al.}~\cite{wang2009a,wang2009b} It is rather similar to the force-fitting approach that we have used for polarizable force field. The idea is to determine effective pairwise forces between coarse-grained sites by averaging over the atomistic forces between the corresponding atomic groups in configurations extracted from all-atom MD simulations (in principle it would also be possible to use AIMD). The size of the grains is smaller than in the models proposed by Roy {\it et al.}, and the potential includes bonded interaction terms, so that it should be possible to keep similar parameters for a wide family of ionic liquids. This approach is therefore very promising as a next step for the study of supercapacitors.

\section{Conclusions}

Since its burgeoning, the field of ionic liquids has taken much profit from the use of molecular dynamics. On the one hand, AIMD provides an unique framework for studying the vibrational properties and the mechanisms of chemical reactions. On the other hand, classical MD allows to reach the time and length scales necessary for a deep understanding of their structural, thermodynamic and transport properties, both in the bulk and at interfaces. In the past ten years, such simulations have provided a quantitative understanding of the structure of the ionic liquids, showing the formation of nanosegregated polar/apolar domains, which has strong consequences on the solvation properties of these media. The determination of transport properties  has proven more difficult: conventional, non-polarizable force fields largely underestimate the diffusion coefficients and polarization effects need to be introduced to recover the correct dynamical behavior. In recent years, two approaches have been proposed. The simplest one is to use rescaled charges for the ions, while the other one consists in introducing explicit dipoles (either using Drude oscillators or point dipoles). Although both approaches provide enhanced dynamics, the latter is more accurate. In addition, the corresponding parameters are more transferable from one compound to another, so that this approach should be preferred for systematic studies. Indeed, an ambitious but feasible objective for future studies should be the prediction of the physico-chemical properties of RTILs prior to their synthesis, in order to target the most adapted one for a given application.

Applications, however, often use ionic liquids under special conditions. Indeed, they are widely used as electrolytes in energy storage devices, and it is then necessary to understand their behavior at complex interfaces. The computational cost associated to heterogeneous systems is usually very large, which implies that these studies are currently tackled using coarse-grained force fields. The loss of atomic details implies that comparisons between specific RTILs become more difficult, but these simulations allow for the understanding of problems which are otherwise difficult to address by experiments only. For example, the field of supercapacitors has recently benefitted greatly from molecular simulations, which have provided interpretations for the changes in the electricity storage ability depending on the geometry of the electrodes based on the structure of the ionic liquids. The dynamical aspects are more difficult to handle since even with coarse-grained force fields the computational costs remain high. There are many other fields in materials science in which molecular simulations of RTILs at interfaces should be able to provide very useful input for optimizing devices: electroactuation,~\cite{lee2013b} lubrication,~\cite{mendonca2013a,padua2013a} Li-ion batteries.~\cite{mendezmorales2014a} 

\section*{Acknowledgements}
C\'eline Merlet, Benjamin Rotenberg, Leonardo Siqueira, Barbara Kirchner, Ari Seitsonen and Paul Madden are acknowledged for their crucial contributions to the development of the force fields reported here. I also thank Jos\'e Nu\~no Canongia Lopes, Ag\'ilio P\'adua and Christian Schr\"oder for useful discussions about polarization effects in ionic liquids. Benjamin Rotenberg is responsible for the final form of Figure 1. 


\end{document}